\documentstyle[12pt,fleqn]{article}
\def\beq {\begin{eqnarray}}
\def\eeq {\end{eqnarray}}
\def\beqn {\begin{eqnarray*}}
\def\eeqn {\end{eqnarray*}}
\def\noeqn {\nonumber}
\def\ni {\noindent}
\def\ie{{\it i.e.}}

\def\eg{{\it e.g.}}

\begin{document}

{\small 
\begin{flushright}{TMU-NT940401, hep-ph/9404224 \\April 1994}
\end{flushright}
}

\begin{center}
    {\large \bf Roles of Diquarks\\
    in the Nucleon for the Deep Inelastic Scattering\\and 
the Non-leptonic Weak Transitions\footnote{
talk presented at
 the International Symposium on Spin-Isospin Responses
 and Weak Processes in Hadrons and Nuclei 
 (March 8-10, 1994, Osaka, Japan).}}\\

\vspace{3mm}
{\large Katsuhiko Suzuki}\footnote
{e-mail address : ksuzuki@atlas.phys.metro-u.ac.jp}


\vspace{2mm}
        {\it Department of Physics, Tokyo Metropolitan University}\\
        {\it Hachiohji,Tokyo 192, Japan}
\end{center}

\vspace{5mm}

\ni
{\bf Abstract} : We study roles of quark-quark correlations in the 
baryons for the deep inelastic structure function and the 
${\Delta}I=1/2$ rule of the non-leptonic weak hyperon decay.  
The quark-quark correlation is incorporated phenomenologically as 
diquarks within the Nambu and Jona-Lasinio model. 
The strong diquark correlations in the spin-0 channel 
enhance the ${\Delta}I=1/2$ weak matrix elements, and 
account for the ${\Delta}I=1/2$ rule.  
The ratio of the nucleon 
structure functions $F_2^{n}(x)/F_2^{p}(x)$ also comes to 
agree with experiment due to the diquark correlations. 
\vspace{1.5cm}

\ni
{{\bf 1 Introduction $-$ quark-quark correlations inside the nucleon}}

Recently models of the baryons have attracted 
considerable interest.  These models, {\eg}  MIT bag model and 
Skyrmion, possess some essential aspects of the low energy QCD, and 
give successful descriptions for static baryon properties such as 
masses and magnetic moments.  However, these models do not take into 
account {\it quark-quark correlations} inside the baryon.  
In the QCD sum rule calculation, asymmetric momentum distributions in 
the proton are found, where the most part 
of the proton momentum is carried by the u-quark with its spin 
directed along the proton spin and the remaining small part is carried 
by the u-d quarks with combined spin-0 [1].  
Instanton liquid model of the QCD vacuum also shows that the quark 
correlations in the spin-0 and spin-1 channels are completely 
different, due to the large attractive correlation in the spin-0 
channel [2].  
These results indicate that the quark-quark correlation in the spin-0 
channel is quite strong, and is of great importance 
to discuss the baryon structure.  
This strong correlation in the spin-0 channel may be understood as 
a close connection between 
the $qq$ pair in the spin-0 channel 
(\(\bar{\psi}c_{A}i\gamma_{5}\psi_{c}\)) 
and the pion (\(\bar{\psi}i\gamma_{5}\psi\)), through 
a particle-antiparticle symmetry, {\ie} 
Pauli-G\"{u}rsey symmetry [3,4].  
(Here, $c_{A}$ are the $SU(3)$ color antitriplet operators, and 
$\psi_{c}$ the charge conjugation of the quark field.)  
Due to the spontaneous breakdown of chiral symmetry,
the Goldstone pion is a collective state of the QCD vacuum.  
Hence, we expect that the strong correlation exists in the 
corresponding spin-0 $qq$ channel.  

\vspace{-1mm}
In this study, we introduce $"Diqurks"$, {\ie}  correlated two quark 
states inside the baryons, to incorporate the quark-quark 
correlation in the phenomenological way [5].  
Assuming the $SU(6)$ spin-flavor symmetry, the baryon wave 
functions are written as products of diquarks and quarks.  
For example, the proton wave function is given by
%
%
\beq
|p\uparrow >&=&{1 \over {\sqrt {18}}}[ 3S(ud)u\uparrow + 
 2A(uu)^{+}d\downarrow 
-\sqrt 2A(uu)^{0}d\uparrow \nonumber\\
& & \hspace {5cm} -\sqrt 2A(ud)^{+}u\downarrow +  
A(ud)^{0}u\uparrow ] \; ,
\label{WF}
\eeq
\noindent
where \(S(ud)\) denotes the scalar (spin-0; $0^{+}$), isospin-0 
diquark, and \(A(ij)^\kappa\) the 
axial-vector (spin-1; $1^{+}$) diquark with the flavor content 
$i$ and $j$, and the helicity $\kappa$.  These quantum numbers are 
determined by the 
antisymmetrization of the diquark wave function.  In order to 
incorporate the non-perturbative quark-quark correlations, 
namely, structures of 
diquarks, we use the Nambu and Jona-Lasinio (NJL) model [6].  
This model demonstrates the spontaneous chiral symmetry breaking, and 
reproduces the $SU(3)_{f}$ meson properties very well with the 
parameters fixed by the pion and kaon properties [3].  
The constituent quark mass, 
the diquark mass, and the diquark-quark coupling constant are obtained 
by solving the Gap equation and the Bethe-Salpeter equation 
simultaneously.  
\vspace {0.5cm}

\ni
{{\bf 2 ${\Delta}I=1/2$ non-leptonic hyperon decay and diquark 
correlation}}

The puzzle of the non-leptonic weak decay and the ${\Delta}I=1/2$ rule 
has a long history.  
In the standard factorization approximation, 
theoretical calculations show serious disagreements with experiments
(Table 1).  Although the renormalization group method for 
gluon exchange contributions indicates an enhancement 
of ${\Delta}I=1/2$ amplitude and a 
suppression of ${\Delta}I=3/2$ amplitude, this effect is too small.

\vspace{-1mm}
Recently, Stech and his collaborators pointed out that 
the scalar diquarks play 
a crucial role for the parity conserving part (P-wave) of 
${\Delta}I=1/2$ transitions [7].  We consider a possible intermediate 
state contribution, which is the so called pole diagram as shown 
in Fig.1, 
where an initial hyperon $B_i$ changes to an intermediate state 
baryon $B_m$ by the weak interaction, and then $B_m$ emits a pion to 
produce a final state baryon $B_f$.  
The scalar diquark appears in the calculation of these weak transition 
matrix elements, since the effective weak 
Hamiltonian is rewritten by the diquark annihilation and creation 
operators by virtue of the Fierz transformation[7].  
%
%
\beq
{\cal H}&=&{{G_{F}\sin \theta \cos \theta } \over {\sqrt 2}} \{
c_{1}[\bar u\gamma _\mu (1-\gamma _5)s]
[\bar d\gamma ^\mu (1-\gamma _5)u]\noeqn\\
& & \hspace {5cm} +c_{2}[\bar u\gamma _\mu (1-\gamma _5)u]
[\bar d\gamma ^\mu (1-\gamma _5)s]  + .... \noeqn\\
&=&  {{G_{F}\sin \theta \cos \theta } \over {\sqrt 2}}
(c_{1} - c_{2}) \frac {2} {3} [\bar u_{c}c_{A}(1-\gamma _5)d]^{\dagger}
[\bar u_{c} c_{A} (1-\gamma _5)s] + ... 
\label{weak}
\eeq
\ni
Hence, this weak process occurs as a transition between spin-0 
diquarks, 
$(us)^{0+} \rightarrow (ud)^{0+}$ with the remaining quark unchanged.  
Note that $(ud)^{0+}$ scalar diquark necessarily has 
$I=0$, and $(us)^{0+}$ has $I=1/2$ due to the Pauli principle.  
Therefore, the strong scalar diquark correlations enhance only the 
${\Delta}I=1/2$ weak matrix elements, and thus explain 
the ${\Delta}I=1/2$ rule.  Using the NJL model to calculate the diquark 
weak decay matrix elements, we obtain numerical results in Table 1 [8].  
The results of the factorization including the penguin contributions 
are listed in the second column, and the diquark contributions in the 
third column.  In the forth column, the sum of the diquark and the 
factorization is listed, which is compared with experimental data 
indicated in the fifth column.  The diquark process enhances largely the 
transition amplitudes, and the result is in a good agreement with 
experiment.  

\vspace {0.5cm}
\hspace{-2mm}{\bf Table 1} P-wave amplitude ($\times 10^{-7}$) 

\begin{tabular}{lcccc}
& & \\ \hline 
& Fact. & Diquark & Total & Exp \\ \hline
${\Sigma}^{+}_{+}$ & 0.00 &  37.0 & 37.0 & 41.8 \\ 
${\Sigma}^{+}_{0}$& 2.05 & 24.5 & 26.5 & 26.7 \\ 
${\Sigma}^{-}_{-}$& $-3.92$ & 2.47& $-1.45$ & $-1.44$ \\ 
${\Lambda}^0_{-}$& 9.60 & 12.0 & 21.6 & 22.4 \\ 
${\Xi}_{-}^{-}$& $-2.40$ & 18.8 & 16.4 & 17.5 \\   
${\Omega}_{K}$ & 0.00 & 6.59 & 6.59 & 5.37 \\ \hline
\end{tabular}

\vspace{0.9cm}
We note that estimations of the pole diagram within other effective 
models of the baryon yield much smaller values than experimental data, 
due to the absence of the quark correlation in those models.  
For instance, the MIT bag model gives 25.3 for ${\Sigma}^{+}_{+}$ 
amplitude, which disagrees with data.  
\vspace {5mm}

\ni
{\bf 3 Deep Inelastic Nucleon Structure Function and Diquarks}

On the other hand, the nucleon structure functions provide us with the 
information on the spin-flavor structure in the nucleon.  The 
ratio of nucleon structure functions $F_2^{n}(x)/F_2^{p}(x)$ shows a 
clear deviation from the naive quark-parton model value; 2/3.  
We evaluate the nucleon structure function within the diquark-quark 
model, and discuss how the flavor structure of the structure function 
depends on the diquark correlation.  

\vspace{-1mm}
In this model, both quark and diquark scattering processes contribute 
to the structure functions [9].  The quark scattering process (Fig.2a), 
in which a quark is struck out by the virtual photon with the residual 
diquark being a spectator, is evaluated by using the standard method in 
the impulse approximation [10].  For the diquark part (Fig.2b), 
we obtain the distribution function as the convolution of quark 
distributions in diquarks 
with the diquark distributions in the nucleon.  This diagram represents 
the leading-twist contributions of diquarks, where diquarks break up 
completely after absorbing the virtual photon.  Concerning the quark 
distributions in diquarks, we use the same procedure 
as done in the meson case in terms of the NJL model [11].

As a result, the proton and neutron structure functions are 
written as
%
%
\beq
F_{2}^{p}(x)&=&{2 \over 9}q^{S}(x)+{1 \over 9}q^{V}(x)+
{5 \over {18}}Q_{D}^{S}(x)+{7 \over {18}} Q_{D}^{V}(x)\noeqn\\
F_{2}^{n}(x)&=&{1 \over {18}}q^{S}(x)+{1 \over 6}q^{V}(x)+
{5 \over {18}}Q_{D}^{S}(x) +{1 \over 6}Q_{D}^{V}(x) \; ,
\eeq
\ni
where \(q^S\) and \(q^V\) are the quark distributions with the residual 
diquarks being the scalar and the axial-vector diquarks, respectively 
(Fig.2a).  
\(Q_D^S\) and \(Q_D^V\) are the quark distributions obtained by the 
scalar and the axial-vector diquark scattering processes, 
respectively (Fig.2b).  The flavor dependence of the structure 
functions arises 
from the difference of the distribution contents in this model.
The correlations in the spin-0 and spin-1 diquarks affect the quark 
distributions, mainly through their masses.  We note that this approach 
naturally reproduces the asymmetry of momentum distributions in the 
nucleon obtained by the QCD sum rule [1].  To see this point, 
we calculate the momentum fraction carried by each quark.  
In the scalar channel, we find  
$<xq^{S}> : <xQ_D^{S}>\sim  2 : 1$, which is consistent with 
the result of ref. [1], while we get 
$<xq^{V}> : <xQ_D^V>\sim 1 : 1$ in the axial-vector channel.  
These results reflect the strength of the correlation in each 
channel.   In the axial-vector channel, the quark correlation is so 
weak that the asymmetry of the momentum distribution is negligible.  

\vspace{-1mm}
Our calculated quark distribution gives a 
boundary condition for the QCD perturbation at the low energy model 
scale $\mu ^2 \sim 1GeV^2$, at which the effective quark model is 
supposed to work [12].  The distribution functions 
are evolved to the experimental high momentum scale with the help 
of the Altarelli-Parisi equation.  

\vspace{-1mm}
The resulting ratio $F_2^{n}(x)/F_2^{p}(x)$ is shown in Fig.2.  
The ratio $F_2^{n}(x)/F_2^{p}(x)$ 
depends on the strength of the quark-quark correlation, and is the same 
value as the parton model prediction without the spin-0 
diquark correlations (dashed curve).  
If the correlation in the scalar channel is 
so strong as to reproduce the non-leptonic transitions, it is in a 
reasonable agreement with the data (solid curve), 
except for the small $x$ region, where the sea quark effects dominate.

\vspace {1.5cm}
\ni
{\bf 4 Summary}

We have studied the diquark correlation inside the baryon, and its 
contribution to the non-leptonic hyperon weak decay and the deep 
inelastic scattering.  The diquark correlation, suggested by the QCD 
sum rule and the instanton liquid model, is the strongest in the spin-0 
and isospin-0 channel.  We comment on the result of the lattice QCD 
calculation, in which the diquark-like clustering is not observed [13].  
At the present, such lattice simulations are performed within the 
quenched approximation.  However, it is essential to include the light 
dynamical quark for the discussion of the hadron structure.  Thus, the 
quenched approximation may be inadequate to study such diquark clusters.  

We have used the simple $SU(6)$ diquark-quark model to incorporate the 
quark correlations within the NJL model.  Such a diquark 
correlation enhances the ${\Delta}I=1/2$ weak transition amplitudes, 
and reproduces the experimental data very well.  
The diquark correlations also produce the asymmetric momentum 
distribution inside the nucleon, as obtained in the QCD sum rule 
calculation.  Due to this asymmetric distribution, the ratio of the 
nucleon structure functions becomes smaller than the naive parton model 
value at the large Bjorken $x$, and is consistent with experiment.  
Our results indicate the importance of 
the quark-quark correlation in the nucleon.  

The author wishes to thank Prof. H. Toki and Dr. T. Shigetani 
for their collaborations in this work.  

{\small

\vskip 2mm
%
%
{\setlength{\parindent}{0pt}{\bf References}}
\newcounter{capn}
\begin{list}{[\arabic{capn}]}
{\usecounter{capn}\setlength{\leftmargin}{1.cm}
  \setlength{\rightmargin}{0cm}}
\vspace{-4mm}
\item{V.L. Chernyak and A.R. Zhitnitsky, Phys. Rep. {\bf 112} 
(1984) 173} 
\vspace{-2mm}
\item{T. Schafer, E.V. Shuryak and J.J.M. Verbaarschot, 
Nucl. Phys. {\bf B412} (1994) 143} 
\vspace{-2mm}
\item{U. Vogl and W. Weise, Prog. Part. Nucl. Phys. {\bf 27} 
(1991) 195, and references therein} 
\vspace{-2mm}
\item{K. Suzuki and H. Toki, Mod. Phys. Lett. {\bf A7} 
(1992) 2867} 
\vspace{-2mm}
\item{For a review, M. Anselmino {\it et al.}, Rev. Mod. Phys. {\bf 65} 
(1993) 1199}
\vspace{-2mm}
\item{Y. Nambu and G. Jona-Lasinio, Phys. Rev. {\bf 122} 
(1961) 345} 
\vspace{-2mm}
\item{H.G. Dosch, M. Jamin and B. Stech, Z. Phys. {\bf C42} (1989) 167} 
\vspace{-2mm}
\item{K. Suzuki and H. Toki, Mod. Phys. Lett. {\bf A}, in press}
\vspace{-2mm}
\item{K. Suzuki, T. Shigetani, H. Toki, hep-ph/9310266, 
Nucl. Phys. {\bf A} in press}
\vspace{-2mm}
\item{H. Meyer and P.J. Mulders, Nucl. Phys. {\bf A528} 
(1991) 589}
\vspace{-2mm}
\item{T. Shigetani, K. Suzuki, H. Toki, Phys. Lett. {\bf B308} 
(1993) 383; TMU preprint (1994), hep-ph/9402277} 
\vspace{-2mm}
\item{R.L. Jaffe and G.G. Ross, Phys. Lett. {\bf B93} 
(1980) 313} 
\vspace{-2mm}
\item{D.B. Leinweber, Phys. Rev. {\bf D47} (1993) 5090} 
\end{list}
}
\newpage
\ni
{\bf Figure Captions}

\vspace{1cm}
\ni
Fig.1 : Pole diagram for the non-leptonic hyperon decay

\vspace{1cm}
\ni
Fig.2 : The forward scattering amplitude of the nucleon.   
The thin solid line represents the quark, and the shaded line 
the diquark.  The nucleon and the virtual photon are depicted by the 
solid and the wavy lines.  

\vspace{1cm}
\ni
Fig.3 : The ratio of the neutron-proton structure functions at $Q^2 = 
15GeV^2$.  The experimental data are taken from the EMC, BCDMS, and 
NMC experiments.  The solid curve is obtained by taking into account the 
strong scalar diquark correlation.  
The dashed curve is the result where the channel difference of the 
diquark mass is neglected [9].

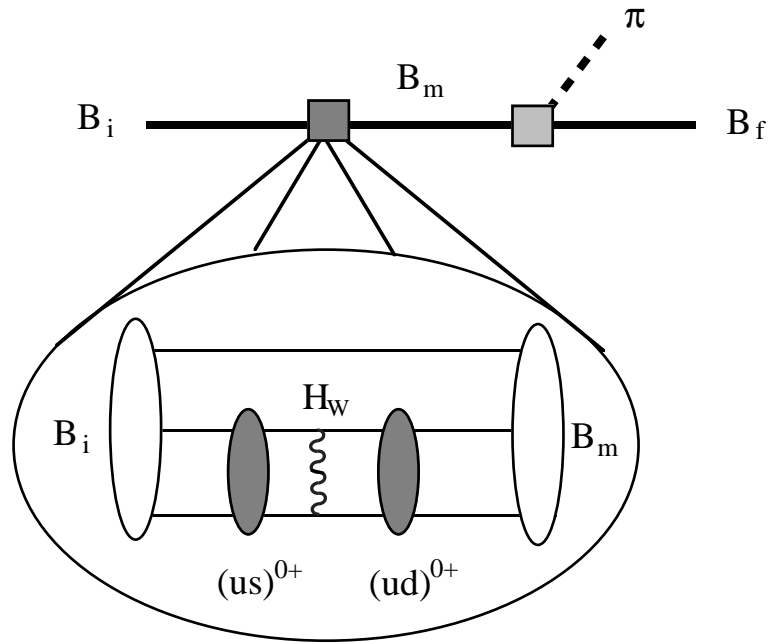
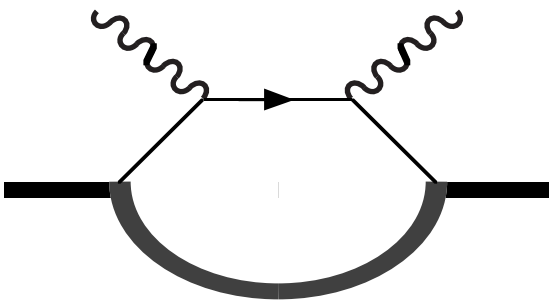
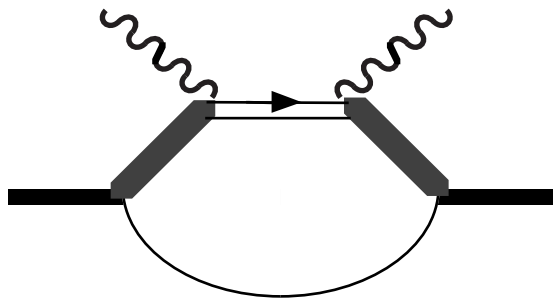
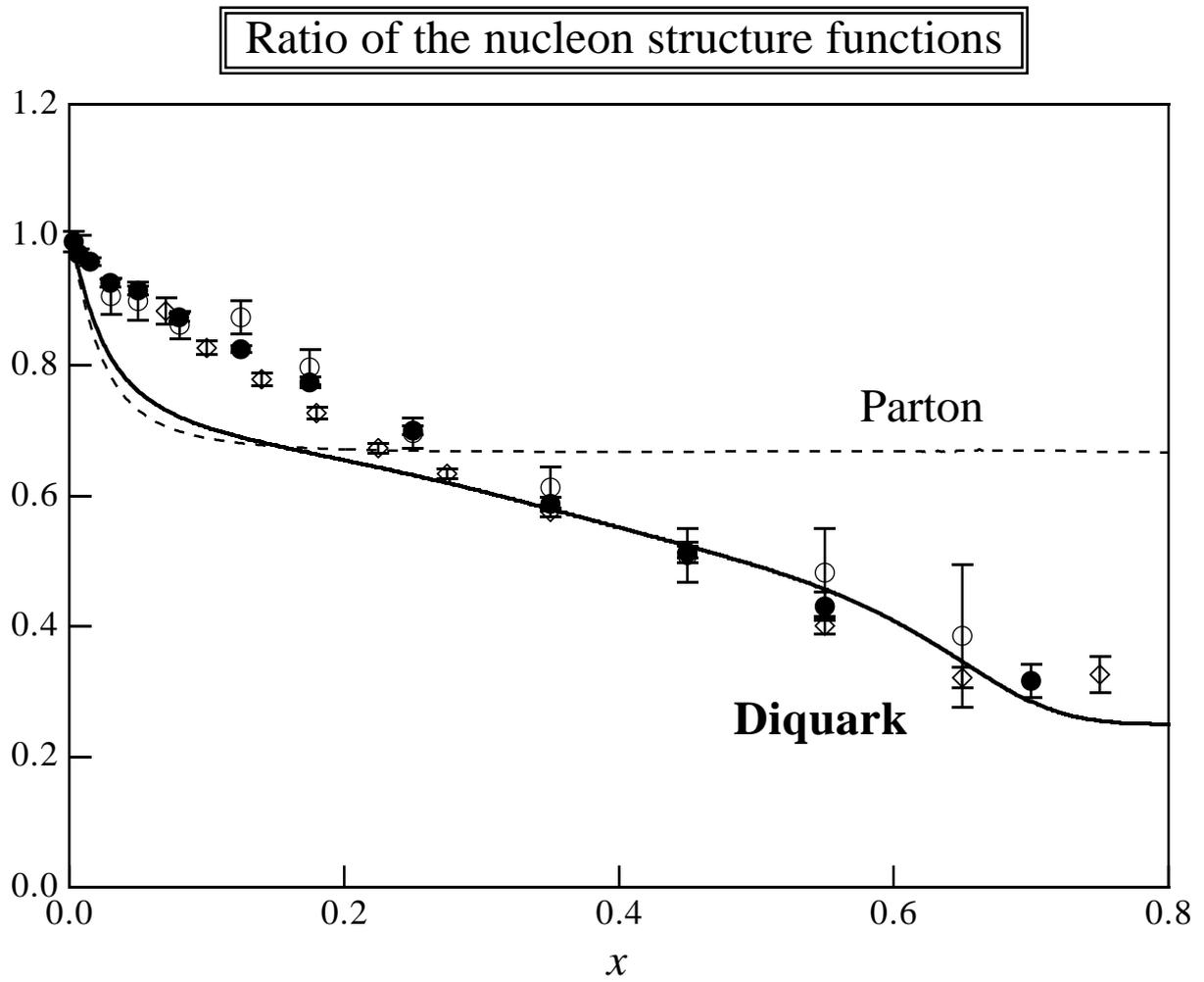
\end{document}